\documentclass[prl,amsmath,amssymb,reprint]{revtex4-1}
\usepackage{graphicx}  
\usepackage{dcolumn}   
\usepackage{bm}        

\usepackage[utf8]{inputenc}
\usepackage{CJK}

\begin{document}
\begin{CJK}{UTF8}{gbsn}

\title{Dipolar Dimer Liquid}
\author{Junyi Zhang}
\email{junyiz@princeton.edu}
\affiliation{Department of Physics, Princeton University, Princeton, 08544, New Jersey, USA }

\date{\today}

\begin{abstract}
A model of dipolar dimer liquid (DDL) on a two-dimensional lattice has been proposed. 
We found that at high density and low temperature, it has a partially ordered phase which we called glacia phase. The glacia phase transition can be understood by mapping the DDL to an annealed Ising model on random graphs. In the high density limit the critical temperature obtained by the Monte Carlo simulation is $k_BT_c^G = (3.5\pm0.1)J$, which agrees with the estimations of the upper and lower bounds of $k_BT_c^G$ with exactly solved Ising models.
In the high density and low temperature limit, we further studied configurational entropy of the DDL in the presence of the neutral polymers.  The suppression of the configurational entropy scales as a power law of the polymer length $\lambda_p^\alpha$ with $\alpha \geq 1$, which implies that the configurational entropy of water plays essential roles in understanding the hydrophobic effect and the protein folding problem.
\end{abstract}

\maketitle
\end{CJK}

\emph{Introduction.}--
Dimers on lattices have been constant topics of statistical physics~\cite{Kasteleyn1961,Temperley1961,Fisher1961,Kasteleyn1963,Lieb1967,Alet2005,Dickman2012,Otsuka2011} and mathematics~\cite{Thurston1990,Elkies1991}.
They have been used for understanding various phenomena, e.g., surface adsorptions~\cite{Retter1987}, liquid-vapor transitions~\cite{Lebowitz1971,Dickman2012}, liquid crystals~\cite{Heilmann1979,Abraham1980}, superconductivity~\cite{Rokhsar1988}, roughening transitions~\cite{Blote1982,Levitov1990}, etc.
Nevertheless they are also simple enough so that they are exactly solvable by various mathods~\cite{Kasteleyn1961,Temperley1961,Fisher1961,Kasteleyn1963,Lieb1967,Elkies1991}.

It is natural to consider the dimers carrying dipole moments for better describing the molecules of the polar solvent, like water ($H_2O$) and hydrofluride ($HF$).  
In addition to the hard-core repulsion, the leading term of the interactions is the electrostatic Coulomb forces due to the polarized charges. (In chemistry, this effect is often referred to as hydrogen bond.)

In this article, we considered a model of dipolar dimers with electrostatic interactions, which we called dipolar dimer liquid (DDL). We found that it has two liquid phases.  In the high density and low temperature phase, the charges of the dimers are ordered while no translational symmetry is broken for the dimer configurations, which we called {\sl glacia} phase.
The phase transition may be understood by mapping the the DDL to an annealed Ising model on random graphs.
In the high density and low temperature limit, the system has massively degenerated ground states  while the charge degrees of freedom are completely frozen.
Therefore it allows to calculate the ground-state entropy exactly~\cite{Kasteleyn1961,Lieb1967,Tiliere1900s,Jacobsen2016}.

A particularly interesting property of water is the hydrophobic effect.  In the text books, a famous rule of thumb in chemistry, ``like dissolves like'', is usually attributed to the disruption of the hydrogen bonds  when a nonpolar molecule is present in the water (enthalpy effect).  A well known example is that oil does not like to dissolve in water.   
The hydrophobic effect has also been recognized as a key ingredient for understanding the protein folding problem which remains an important open problem in biochemistry over decades.  
However, it may not be properly understood without including the entropy~\cite{Silverstein1998}.  
The entropy contributions come from both the conformations of the proteins and the configurations of the solvent~\cite{KAUZMANN1959,BRADY1997,schafer2001,Miyamoto1993}.
Much effort has been made on conformation entropy of the proteins~\cite{Doig1995,Zhang2006,Baxa2014,Chang2007} but fewer on the configurations of the solvent~\cite{Breiten2013}.

We calculated the configurational entropy of the DDL in the high density and low temperature limit in the presence of neutral polymers of different conformations.  Surprisingly, we found the suppression of the configurational entropy of the DDL depends strongly on the polymer conformations.
The dimers surrounding the polymers are more ``ordered'' although there are no liquid-solid phase transitions.
It provides a quantitative understanding of the picture of forming water ``cages'' around the polymers in the theory of hydrophobic effect.

The DDL is undoubtedly an over-simplified model for the realistic polar solvent.  Nevertheless, in such a simple model, it is much easier to identify and distinguish various effects observed in more complicated systems, e.g., the proteins in the water.
As one may map the DDL to the Ising models, and it is exactly solvable in some limits, which provides a bench mark for more realistic models and numerical simulations. 

\emph{Definition of models.}--
Let  $L = \{V, E\}$ be a two-dimensional lattice, where $V$ and $E$ are vertex set and edge set respectively. 
Each dimer sits on an edge $e \in E$ of the lattice and it covers two neighboring vertices $v_1,v_2 \in V$ linked by the edge $e$.
Each dimer is charge neutral in total but its two ends carry an equal amount of charges but of opposite signs, therefore the dimer has a dipole moment $d_e = Q l_e$, where $l_e$ is the length of the dimer.
The dimers are not allowed to overlap on top of each other and they may also have electric interactions due to the polarized charges and dipole moments. 
This serves a very general definition of the DDL. In order to capture some characters of the realistic systems and to be mathematically handleable, we shall specify some features in more details.

Since the microscopic molecules are identical, it is natural to assume that the lattice holding the dimers has its edges of the same length $l_0$.  We shall also assume that the lattice is bipartite, i.e., there exist two subsets $A,B$ of $V$ such that $A \cup B = V, A \cap B = \emptyset$, and that for each edge $e$, it has one vertex in $A$ and the other in $B$.  Particularly, in this article we shall focus on the case of the two dimensional square lattice.
Denote $C$ the set of the edges covered by a dimer for a given configuration of dimer-covering.
The molecules of the polar solvents have permanent dipole moments. One assumes that each dimer carries a dipole moment of the magnitude $d_0$, therefore $Q_0 = d_0/l_0$ is the amount of the charges that each vertex of a dimer has.

The leading term of the interactions is the Coulomb interaction.  For each $v$ vertex covered by a dimer it has a charge $Q(v) = \pm Q_0$; otherwise it has no charge $Q(v) = 0$.   We neglect the long-range tail of the Coulomb potential and only consider the nearest neighbor interaction.  So the Hamiltonian for the Coulomb interaction is
\begin{equation}\label{eq:Model_CoulombHam}
\begin{split}
H_Q =& \sum_{e\in E; v_1,v_2 \in \partial e} J(e|C) Q(v_1) Q(v_2),\\
J(e|C) =& 
\begin{cases}
+\infty &\text{ for  $e\in C$,}\\
J>0 &\text{ for $e \not\in C$,}
\end{cases}
\end{split}
\end{equation} 
where $J$ is a positive coupling constant.

This term reflects the partially-covalent nature of the hydrogen bond.   The number of the hydrogen bond an electronegative atom can have is determined by its alone pairs, e.g., $O$ in $H_2O$ may form two hydrogen bonds and $F$ in $HF$ may form three.  For the dimers on the square lattice, an atom on a vertex is covalently bonded to one of its neighbors, and it can form hydrogen bond with three more unbonded neighbors.
To take the angled geometry of the water molecules into account, one may improve the model by putting angled trimers on a Lieb lattice
~\footnote{The oxygen atom sits on a the degree-four vertex and its two orthogonal pods lie on the neighboring edge centers.}, which has been reported experimentally in Ref.~\citenum{AlgaraSiller2015}

The subleading electrostatic interaction is the dipole-dipole interaction
\begin{equation}\label{eq:Model_DipoleHam}
\begin{split}
H_D = \frac{\mathbf{d}_1 \cdot \mathbf{d}_2 - 3(\mathbf{d}_1 \cdot \mathbf{e}_r)(\mathbf{d}_1 \cdot \mathbf{e}_r) }{r^3},
\end{split}
\end{equation}
where $\mathbf{r}$ is the vector from one dipole to the other and $\mathbf{e}_r$ is the unit vector along the direction of $\mathbf{r}$.
The dipole interaction tends to align two dimers in some preferred directions relative to each other.
Some lattice dimer models with neutral dimers and direction-dependent interactions have been discussed by Heilmann and Lieb~\cite{Heilmann1979} for describing  the liquid crystals.
In Ref.~\citenum{Heilmann1979}, it has been proved that at high enough fugacity and at low enough temperature there is a nematic phase.
In the simple solvents like water or hydrogen fluoride, there is no observed evidence of the existence of such a nematic phase, so we neglect the dipole-dipole interaction.

One may also take the ionization effect into account by introducing charged monomers or trimers ($OH^-$, $H_3O^+$ for water and $HF_2^{-}$, $H_2F^+$ for hydrogen fluoride).
As $pK_w = 14$~\footnote{$K_w$ is the dissociation constant of the water self-ionization.} around the room temperature, the fraction of the ionized water molecules is negligible.
When the salts, like the strong electrolyte $NaCl$, are added, the density of the ions and cations in the solution is no longer negligible.  It is then necessary to consider the charged monomer-dimer systems instead of the simple DDL.

Therefore we shall consider a DDL on a square lattice $L$ with the short range Coulomb interaction given by Eq.~\ref{eq:Model_CoulombHam}.  Although it is such a simplified model compared to the real systems, we believe that it in some sense reasonably approximates the polar solvents.  On the other hand, this model is simple enough so that we may solve it exactly in some limits.  We will show that the liquid has two phases.  In the high density and low temperature limit, it reduces to a dimer-covering problem, which allows us to calculate the configurational entropy even in the presence of a neutral polymer.

\emph{Phases of dipolar dimer liquid.}--
The dimer liquid defined in the previous section has three parameters, the coupling constant $J$, the temperature $T$ and the density of the dimers $\rho$.

When the dimer density $\rho$ is small,  the nearest-neighbor Coulomb interaction between the dimers is negligible and the no-overlap constraint can be considered as a short range hard core repulsion.  So the system is in the vapor phase as described by the Widom-Rowlinson model~\cite{Widom1970}.  The existence of a liquid-vapor phase transition has been proved in the perspective of grand canonical ensemble by Lebowitz and Gallavotti~\cite{Lebowitz1971} as well as Ruelle~\cite{Ruelle1971} by using a Peierls contour argument.
In the high density limit the system is in the liquid phase.

In the high density limit, the existence of a charge order phase transition can be made in analogy to the 
model of two-component Widom-Rowlinson mixture.~\footnote{This picture was pointed to the author by Elliott Lieb.}
A vertex $v\in V$ of the square lattice $L$ is labelled by its row-column coordinate $(m,n)$, we defined its signature $\sigma(v) = (-1)^{(m+n)}$, which is consistent with $A$-$B$ partition of the lattice.  If a dimer with its positive charge on a vertex in set $A$,  we call it a dimer of $A$-type and color it in red;
if a dimer with its positive charge on a vertex in set $B$,  we call it a dimer of $B$-type and color it in blue.  As the lattice $L$ is bipartite, any dime must be either $A$-type or $B$-type. 

We call two dimers are neighbors, if there exists an edge $e$ whose two ends are one vertex of each dimer respectively.  There are four kinds of neighbors: 1) a pair of collinear dimers, 2)  a pair of perpendicular dimers, 3) a pair of parallel dimers not on the same square plaquet, 4) a pair of parallel dimers sitting on the opposite edges of one sqaure plaquet.
Then the dimers of the same type in a configuration of neighbor kind 1)--3) attract each other with strength  $J$, and in a configuration of neighbor kind 4) with $2J$; the dimers of different types in a configuration of neighbor kind 1)--3) or neighbor kind 4) repel  each other with $J$ or $2J$ respectively.
The no-overlap constraint leads to a hard core repulsion.

The order parameter is defined as the difference of the densities of red and blue dimers
\begin{equation}\label{eq:Phase_OrderParam}
\begin{split}
\Phi = \rho_r - \rho_b
\end{split}
\end{equation} 
It follows the Peierls contour argument~\cite{Lebowitz1971,Ruelle1971} that there is a red-blue symmetry broken phase when the fugacities $z_A =z_B$ is high enough and the dimensionless coupling $\beta J$ is strong enough.  We call this phase glacia phase.~\footnote{{\sl Glacia} means ice in Latin.  In the glacia phase, the charge of the dimers are ordered, while no translational symmetry is broken for the dimer configurations.  Mapping the DDL to the Ising model shows the system looks similar to a glass model.}

An more quantitative way to understand this phase transition is to map the DDL to an \emph{annealed Ising model on random graphs}.
Consider the Lieb lattice $\tilde{L}$ associated to the square lattice $L$, i.e., adding a site $\tilde{v}_e$ in the middle of each edge $e$.  For a dimer of type $A$ on a edge $e$, we associate a spin pointing upwards to the  Lieb site $\tilde{v}_e$, and for type $B$, we associate a spin  pointing downwards to $\tilde{v}_e$.
The no-overlap constraint of the dimers is interpreted as no two spins are put on the nearest-neighboring  Lieb sites.
For two neighboring dimers of kind 1)--3), one links  the corresponding Lieb sites with a single bond; 
for two neighboring dimers of kind 4), one links  the corresponding Lieb sites with a double bond.
For a given configuration of the dipolar dimers, the Lieb sites assigned with a spin and the bonds linking the Lieb sites form a  random graph $G = \{\tilde{V}^L, \tilde{E}\}$.
For each bond, we assign an spin-spin interaction $(-J)\sigma_{\tilde{v}_1} \sigma_{\tilde{v}_2} $. (For a double bond, one counts the bond multiplicity, i.e., in total $2J$ for the spins connected by the double bond). 
The partition function is
\begin{equation}\label{eq:Model_PartitionFunction}
\begin{split}
Z =& \sum_{C} \sum_{\{\sigma\}|C} \mathrm{e}^{-\beta H_\sigma} = \sum_{C}  Z_C,\\
H_\sigma =& \sum_{b\in \tilde{E}; \tilde{v}_1,\tilde{v} _2 \in \partial b} (-J)\sigma_{\tilde{v}_1} \sigma_{\tilde{v}_2} .
\end{split}
\end{equation} 
In general, one may add a weight factor $W_C$ before the partition function $Z_C$ of the Ising model on the random graph $G_C$ associated to the dimer configuration $C$.  
The partition function $Z_C$ is averaged over configurations of dimers, which we call \emph{annealed Ising model on random graphs}.~\footnote{It is a convention in the spin glass that ``annealing'' refers to averaging the partition function over the disorders.}
The order parameter in this picture, i.e., the magnetization for the Ising model
\begin{equation}\label{eq:Phase_OrderParamMag}
\begin{split}
M = \frac{1}{||L||} \sum_{\tilde{v} \in \tilde{V}} \sigma_{\tilde{v}},
\end{split}
\end{equation} 
is the same as the red-blue order parameter $\Phi$ defined for the Widom-Rowlinson mixture.
Instead of the fugacity in the picture of Widom-Rowlinson mixture, it is easier to used the parameter of dimer density $\rho$ in the picture of Ising model.

When the density of the dimers is low, the random graphs are sparse, the spins are not coupled, so there is no spontaneous magnetization.
When the density of the dimers is higher, the spins on the random graphs are more strongly coupled, although there might not be spontaneous magnetization at a finite temperature.
For example, if on average a dimer has less than two neighbors, comparing to the one-dimensional Ising model of which the critical temperature $T_c = 0$,  one would not expect a spontaneous magnetization at a finite temperature.
When the density is even higher, there is a finite critical temperature $T_c$, below which the DDL is in the glacia phase.

The high density limit means all the vertices of the lattice $L$ are occupied by some dimer, i.e., the dimers completely cover the lattice $L$ without any monomer or vacancy.
Then each vertex of a random graph $G_C$ for a given dimer configuration has degree $6$ counting bond multiplicity, although the number of its neighbors may various.~\footnote{A vertex can have: 1) six neighbors all single-bonded, or 2) five neighbors with one double-bonded, or 3) four neighbors with two double-bonded.} 
It is heuristic to compare this annealed Ising model on random graphs to the two dimensional Ising models on the triangular lattice and square lattice, where each site has six or four neighbors but no double bonds.
Each spin on a random graph has at least four neighbors with couplings $J$ or $2J$.  So the critical temperature of the glacia transition should be higher than that of the square lattice, and comparable to that of the triangular lattice $T_c^{\Box}< T_c^G \sim T_c^{\Delta}$.  More precisely, for an excitation of two neighboring spins, on triangular lattice it costs $10J$; on the random graphs, it costs $10J$ or $8J$, which depends on the bond multiplicity. So $T_c^G \lesssim T_c^{\Delta}$.  For some graphs with as many double bonds as possible, it should be comparable with anisotropic Ising model on a square lattice with $J_{h}=J$ and $J_v = 2J$.  The critical temperature is given by $\sinh(2\beta_c^{\Box hv}J_v)\sinh(2\beta_c^{\Box hv}J_h)=1$~\cite{Kramers1941}, which serves as a better estimation of the lower bound of $T_c^G$, i.e., $T_c^{\Box hv}\lesssim T_c^G$.

\begin{figure}[htbp]
\begin{center}
\includegraphics[scale = 0.37]{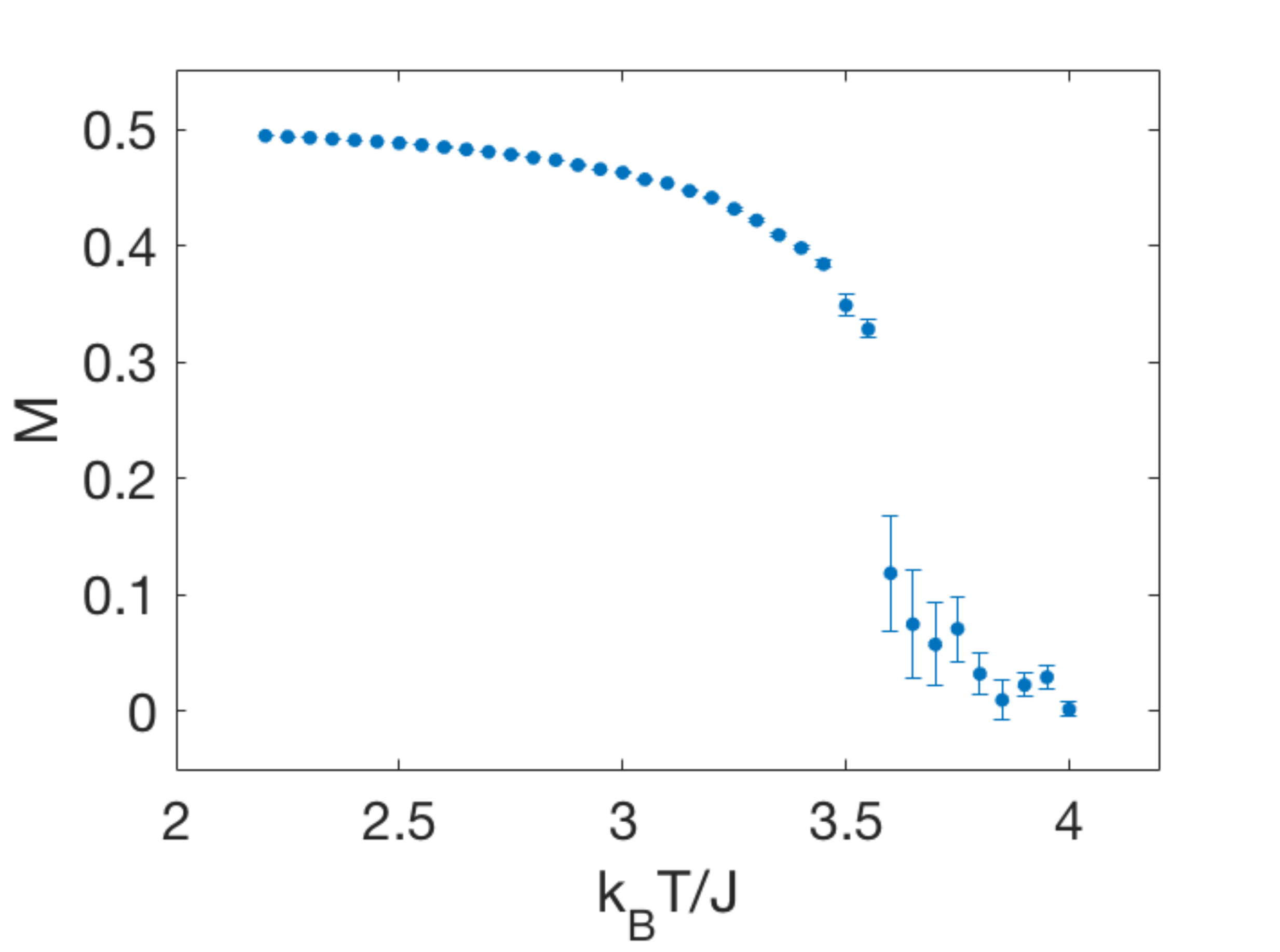}
\caption{\label{fig:Phase_GlaciaTransition} Glacia transition of the DDL at high density limit.  Monte Carlo simulation was done on a $52\times52$ square lattice. The critical temperature  $k_BT_c^G = (3.5\pm0.1)J$.}
\end{center}
\end{figure}

The critical temperatures of the square lattice $k_BT_c^{\Box} = \frac{2J}{\ln(1+\sqrt{2})} \approx 2.2692 J$ for isotropic case and $k_BT_c^{\Box hv} \approx 3.2820 J$ for anisotropic case , and that of the triangular lattice $k_BT_c^{\Delta} = \frac{4J}{\ln 3} \approx 3.6410 J$. 
Fig.~\ref{fig:Phase_GlaciaTransition} shows the Monte Carlo simulation of the glacia transition in the high density limit.
The critical temperature $k_BT_c^G \approx (3.5\pm0.1) J$.

So far we have shown that the DDL has a glacia phase when the dimer density is  high and the temperature is low.  A particularly interesting limit is the high density and low temperature limit, where the dimers completely cover the lattice $L$ and the charges are ordered.
In this limit, as the charge degrees of freedom are frozen, it reduces to the dimer-covering problem which 
is known to be exactly solvable~\cite{Lieb1967,Kasteleyn1961,Kasteleyn1963,Fisher1961,Temperley1961}.  Any dimer-covering configuration admit two ground states, all dimers are of either $A$-type or $B$-type. The ground states are massively degenerate due to the exponentially many possible ways of covering the lattice $L$.   In the thermodynamic limit, the number of arrangements per dimer is $\mathrm{e}^{2G/\pi}$, where $G\approx 0.91596$ is the Catalan constant.
For a given lattice $L$, we define the ground state configurational entropy (not including charge degrees of freedom)
\begin{equation}\label{eq:Phase_ConfigEntropyDef}
\begin{split}
S_G(L) = k_B \ln W(L),
\end{split}
\end{equation}
where $W$ is the number of the ways that dimers completely cover $L$. 

\emph{Configurational entropy in the presence of the polymers.}--
We are also able to calculate the configurational entropy in the presence of the polymers. For polymers with fixed conformations, the configurational entropy is the entropy of the surrounding solvent molecules, which should be distinguished from the conformational entropy of the polymer.
In the thermodynamic equilibrium, it is the total entropy of the system that is maximized, which should include both the conformational entropy of the polymers and the configurational entropy of the solvent.

When the density of the polymers is low, we may consider a system consists of only one polymer surrounded by the DDL.
By definition, a polymer is a string of dimers $D^p = (d^p_1, d_2^p, \dots,  d_N^p)$.  We call the integer $N$ the length of the polymer denoted as $\lambda_p = N$, and we call these dimers constituent dimers.   The constituent dimers do not overlap with each other and any two consecutive dimers $(d^p_i, d^p_{i+1})$ are neighbors.  
The constituent dimers may also carry charges and dipole moments, as the hydrophilic residues of the amino acids or nucleobases of the nucleotides.
A polymer may form loops, where two dimers that are not consecutive in the series $D^p$ sit on neighboring sites and be joint together with a chemical bond, e.g., the disulfide bond in the proteins.  
For simplicity, we shall constraint ourselves to the neutral polymers.

Suppose we fix a conformation $\Omega_p$ of the polymer of length $\lambda_p = N$, and put it in the DDL.
In the high density and low temperature limit, we may calculate the configurational entropy of the DDL
\begin{equation}\label{eq:Entropy_ConfigEntropyDefDDL}
\begin{split}
& S_{DDL}(L|\Omega_p) = S_G(L|\Omega_p) + S_{\text{C}} \\
=& k_B \ln W(L|\Omega_p) + N_D k_B\ln 2,
\end{split}
\end{equation}
where ``$|\Omega_p$'' means in the presence of  a polymer in the given conformation $\Omega_p$,  $W(L|\Omega_p)$ is the number of the ways that the dimers completely covers $L$ in the presence of the polymer, and $N_D$ counts the number of the connected domains, into which the polymer cuts $L$.  
The loss of the entropy due to the presence of the polymer is measured by
\begin{equation}\label{eq:Entropy_ConfigEntropyDefPolymer}
\begin{split}
S_{\text{Config}}(\Omega_p) =& \lim_{L\rightarrow \infty}(S_G(L)  - S_G(L|\Omega_p) )\\
&+ (1 - N_D) k_B\ln 2.
\end{split}
\end{equation}
The limit ${L\rightarrow \infty}$ means the scale of the lattice is much larger than the scale of the polymer, then $S_{\text{Config}}(\Omega_p)$ should be independent of the lattice but only the conformation of the polymer.
By abuse of language, we call  $S_{\text{Config}}(\Omega_p) $ configurational entropy of the polymer for the conformation $\Omega_p$.
$W(L|\Omega_p)$ is counted in a modified Kasteleyn's method~\cite{Jacobsen2016,Tiliere1900s}.

\begin{figure}[htbp]
\begin{center}
\includegraphics[scale = 0.33]{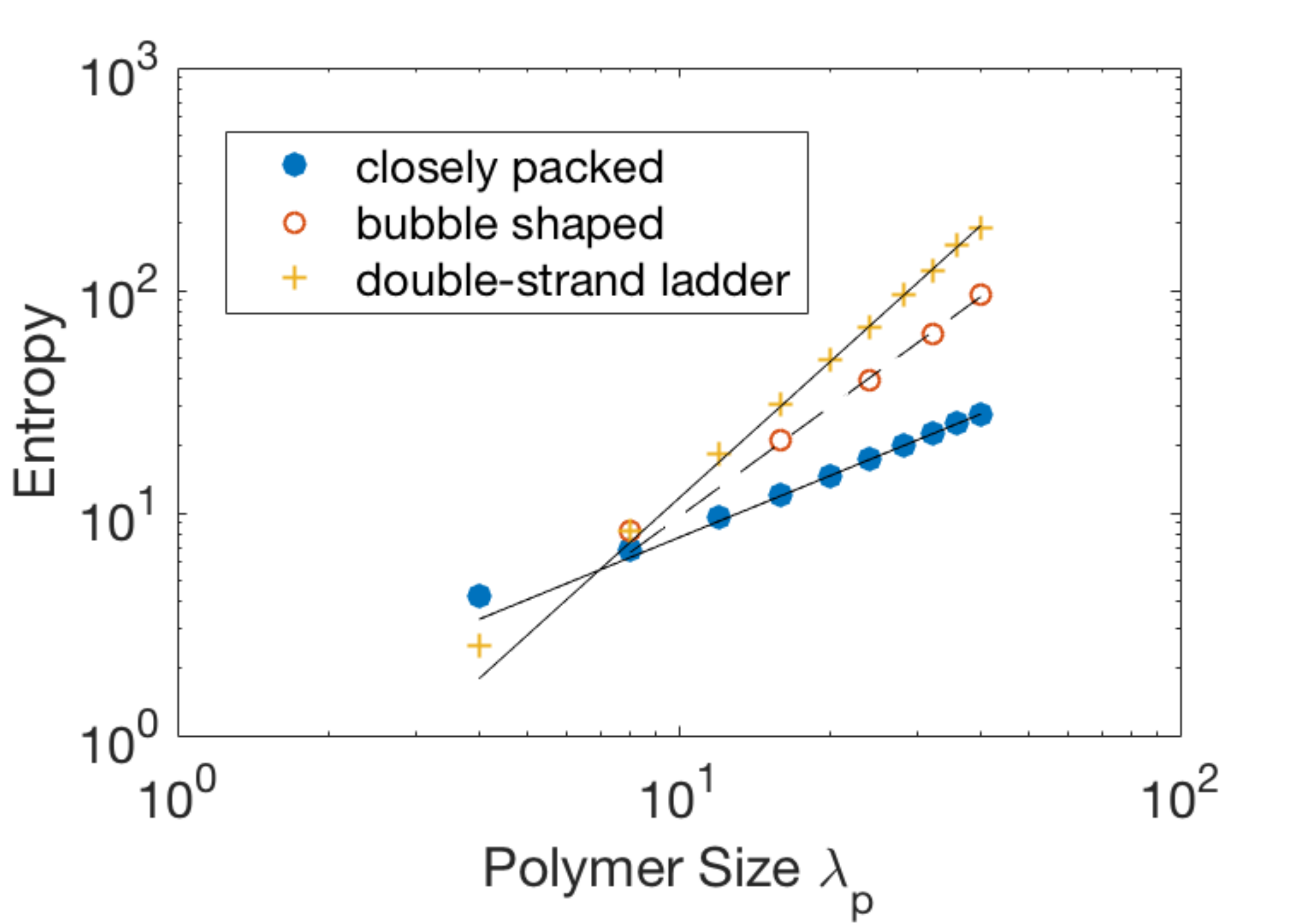}
\caption{\label{fig:Entropy_ConfigEntropy} The configurational entropy of the polymers in different shapes.
Solid dots are $S_{\text{Config}}(\Omega_p = \|)$ for closely packed polymers. 
Circles and crosses are $S_{\text{Config}}(\Omega_p = \#)$ for bubble shaped polymers and $S_{\text{Config}}(\Omega_p = \#)$ for double-strand ladder respectively (with $S_{\text{Config}}(\Omega_p = \|)$ subtracted).  The solid lines scale as $\lambda_p^{0.92}$ (solid dots) and $\lambda_p^{2.0}$ (crosses)  and the dash line  as $\lambda_p^{1.6}$ numerically fitted to the asymptotic entropy scaling.}
\end{center}
\end{figure}
Na\"ively, one would expect $S_{\text{Config}}(\Omega_p)$ is proportional to the length of the polymer $\lambda_p$.  We considered three kinds of conformations: 1) closely packed ($\Omega_p = \|$), 2) double-strand ladder ($\Omega_p = \#$) and 3) bubble shaped ($\Omega_p = O$).
It turns out that this is not always the case.  
Fig.~\ref{fig:Entropy_ConfigEntropy} shows the configurational entropies $S_{\text{Config}}(\|)$, $S_{\text{Config}}(\#) - S_{\text{Config}}(\|)$ and
$S_{\text{Config}}(O) - S_{\text{Config}}(\|)$ in log-log scale.  
For closely packed polymer, the configurational entropy scales linearly with respect to the polymer length $\lambda_p$, which agrees with the na\"ive expectation.
For double-strand ladder polymer, $S_{\text{Config}}(\#) - S_{\text{Config}}(\|)  \propto \lambda_p^2$ asymptotically.
For bubble shaped polymer, $S_{\text{Config}}(O)- S_{\text{Config}}(\|) \propto \lambda_p^\alpha$ with $\alpha=1.6\pm0.1$.~\footnote{Ref.~\citenum{Elkies1991} studied random domino tiling of Aztec diamond, and found a similar effect for the entropy.}

If one examines the microscopic configurations of the dimers, one find that the polymer does not only exclude a certain volume of the liquid, but also aligns partially the dimers  around it.  This seems to be consistent with the empirical picture for the hydrophobic effect in chemistry. The water molecules form the ``cages'' around the polymers.  It is important to note that, since for conformations not closely packed, $S_{\text{Config}}(\Omega_p) \sim \lambda_p^\alpha$ with $\alpha>1$, the configurational entropy growths rapidly with the scale of the polymer. 
In the protein folding problem, it implies that the conformations of the proteins might not be properly understood if the surrounding water molecules were neglected.

\emph{Conclusion.}--We studied the glacia phase transition of the DDL  by mapping it to the annealed Ising model on random graphs.  In the high density limit the critical temperature is estimated to be $k_BT_c^G =(3.5\pm0.1)J$.  We also calculated the configurational entropy of the DDL in the presence of the polymers and found a non-trivial scaling of the suppression of the configurational entropy.  The DDL may serve as a simple model for understanding the complex phenomena in polar solvents.

\begin{acknowledgments}\label{sec:Acknowledgements}
The author thanks Elliott Lieb for helpful discussions particularly for pointing the Widom-Rowlinson picture.  The author also thanks F.~D.~M.~Haldane,  Nozomi Ando, Yimo Han and Guang Chen for motivating discussions.
\end{acknowledgments}

\bibliography{DDLref}
\end{document}